# Planning system for deliveries in Medellín

Catalina Patiño-Forero
Universidad EAFIT
Medellín, Colombia
cpatin10@eafit.edu.co

Mateo Agudelo-Toro
Universidad EAFIT
Medellín, Colombia
magude29@eafit.edu.co

Mauricio Toro
Universidad EAFIT
Medellín, Colombia
mtorobe@eafit.edu.co

## ABSTRACT
Here we present the implementation of an application capable of planning the shortest delivery route in the city of Medellín, Colombia. We discuss the different approaches to this problem which is similar to the famous Traveling Salesman Problem (TSP), but differs in the fact that, in our problem, we can visit each place (or vertex) more than once. Solving this problem is important since it would help people, especially stores with delivering services, to save time and money spent in fuel, because they can plan any route in an efficient way.

To solve this we need to construct a subgraph with the delivering points, based on the city's map, and it will be a complete one i.e. all of its vertices are connected. Then we will give the user different options that will determine which algorithm will be used to solve the problem. Between these options there is only one that will surely give the shortest route and works only with twenty or less points. The other options are quite fast but may or may not give the shortest route.

Depending on the chosen algorithm, the results in time, memory and total distance will vary. For example, we have an algorithm that does not create a subgraph to give an answer, so it takes less memory and time, but will not give the total distance. Others can give a better answer quite fast, even though they require to compute a subgraph, but still the tour produced may not be the shortest one. At last, there is an algorithm that can give the shortest route every time, but needs to look through all possible answers so it takes much more time.

For the problem of planning delivery routes in Medellín our proposed solution to find the shortest route can be of huge help for small companies if their couriers do not visit more than 20 points per trip.

## ACM Classification Keywords
Theory of computation → Design and analysis of algorithms → Approximation algorithms analysis → Routing and network design problems

## 1. INTRODUCTION
Efficiently planning the deliveries is something really useful for any company in the field. Here we talk about creating an efficient program that gives an optimal delivering route for a courier, in order to minimize the time spent traveling; the courier can pass over one place more than once. Without this last condition we would have a TSP which, though it is a "simple" problem formulated over 200 years ago [9], does not have any optimal solution for big graphs (thousands of vertexes). Since it is simpler (and possible) to treat our problem as TSP, we are going to do so.

We will see the different approaches to this problem and also discuss the selection of the best available choice for our specific case.

## 2. PROBLEM
As we just stated, we are trying to create an efficient program that gives an optimal (shortest total distance) delivering route for a courier, which minimizes the time spent traveling; this route can repeat places which were already visited. In our case, we will implement it for the city of Medellín in Colombia, but it does not mean the algorithm cannot be used for other cities.

This efficient route planning request is quite difficult to compute if we want to get an optimal answer. This is due the incredible amount of possibilities we will have, since the idea is to use the algorithm for real cities, for example Medellín, which has a population that surpasses the 2 million people [11]. So it is to be expected that the algorithm will take an incredible amount of time to give an appropriate answer, time that may exceed what we can spend on it. We can take the TSP as an example, which requires a time proportional to $(n-1)!/2$ to execute (where n is the number of places or nodes) [10], which means that for 20 destinations it would require a about 12 years to compute using an average computer. We will treat our problem as TSP but using a faster algorithm that requires less than 3 seconds to compute the path for 20 points, but that would require 14 years for 45 points on the same computer.

## 3. RELATED WORK

### 3.1 Minimum Spanning Tree (MST)
Given a weighted graph, the MST is the cheapest subset of edges that keeps the graph in one connected component [1]. Those are very useful because they give approximate solutions to the traveling salesman problem very efficiently.

One efficient way to compute the MST of a graph is the Kruskal's algorithm. It is a greedy algorithm that starts by placing each vertex on its own connected component. It then iterates over the edges having them sorted in non-descending order, merging the two vertices connected by the current edge if, and only if, they are currently on different components. The complexity of this algorithm is $O(m*\log(m))$ where m is the number of edges.

### 3.2 Hamiltonian Path and Cycle
A Hamiltonian Path is a path between two vertices of a graph that visits each vertex exactly once [5]. A Hamiltonian Cycle is a closed loop through a graph that visits each node exactly once [6]. A closed loop is a cycle in a graph in which the first vertex is the same as the last [7]. A graph possessing a Hamiltonian Path is said to be a Hamiltonian Graph [8].

There is a backtracking approach to find whether an undirected graph is Hamiltonian. We start by creating an empty array and adding vertex 0 to it. We will try to add the other vertices starting from 1, but before that, we check whether it is adjacent to the previously added vertex and if it is not already added. If we find such vertex, we add it as part of the solution. If we do not find a vertex then we return false [1]. Anyway, the complexity of this algorithm is O(n!) where n is the number of vertices, just like the naïve approach.

### 3.3 Eulerian Path and Cycle
An Eulerian Path is a path in a graph that visits each edge exactly once [3], and an Eulerian Cycle is an Eulerian Path which starts and ends in the same vertex [2]. It is similar to the Hamiltonian path because in both we want to visit some part of the graph only once. The difference is that in this case we want to walk through each edge instead of visiting each vertex. This difference changes everything: while the Hamiltonian path is an NP-Complete problem for a general graph, finding whether a given graph is Eulerian (has an Eulerian Cycle) can be done in O(n + m), where n is the number of vertices in the graph and m the number of edges.

To find whether a undirected graph is Eulerian it must have all its non-zero degree vertices connected (which can be done using a DFS traversal) and the number of vertices with odd degree must be 1 (if it is 2 then the graph has a Eulerian Path instead) [4].

### 3.4 Chinese Postman Problem (CPP)
In this problem, given a weighted graph, the postman wants to find the shortest path that visits every edge at least once returning to the starting point.

This problem can be solved in an optimal way by adding the appropriate edges to the graph to make it Eulerian, because that is basically what the problem is: finding an (especial) Eulerian Cycle in the graph. Specifically, we find the shortest path between each pair of odd-degree vertices in the graph. Adding a path between two odd-degree vertices in G turns both of them to even-degree, moving G closer to becoming an Eulerian graph. Finding the best set of shortest paths to add to G reduces to identifying a minimum-weight perfect matching in a special graph G'. For undirected graphs, the vertices of G' correspond the odd-degree vertices of G, with the weight of edge (i, j) defined to be the length of the shortest path from i to j in G. For directed graphs, the vertices of G' correspond to the degree-imbalanced vertices from G, with the bonus that all edges in G' go from out-degree deficient vertices to in-degree deficient ones. Thus, bipartite matching algorithms suffice when G is directed [1]. Once the graph is Eulerian, the actual cycle can be extracted using the algorithm described above.

### 4. DATA STRUCTURES
To implement our algorithm, we need two graphs: one to store the city itself and other to store the points we need to visit.

For the graph of the city, we use adjacency list representation. To achieve this we created four classes: vertex, edge, point and pair. The first two, vertex and edge, are used directly on the representation of the graph: we use HashMaps with vertex objects as keys and edge objects as values. Point class represents the latitude and longitude of the vertex in the world, and is used to access vertices given its latitude and longitude. Pair objects are used during the execution of A* and Dijkstra's algorithm, which require a priority queue of vertices (first value of the pair) sorted by some value acquired during the execution of the algorithm (second value of the pair).

The other is a complete graph that contains only the vertices we want to visit. It is stored as an adjacency matrix, using a primitive double's 2D array with dynamically assigned integer codes to vertices (used as indices), and where the edges are completed using either Dijkstra's or A* algorithms on the city's graph (depending on the amount of points). Since it is a complete graph, using an adjacency matrix is better than an adjacency list, because both need the same memory space, but the adjacency matrix is faster when looking up for the distance between any two vertices and that is a main requirement of our algorithms.

There are other auxiliary data structures used during the execution of different parts of the program. The Natural Approximation algorithm uses a HashMap with vertices as keys and integers as values to remember the original positions of the vertices to visit. To read the vertices the user wants to visit, we use a HashMap with points as keys and vertices as values, because the URL only contains information about the point and not the vertex itself. When reading the edges, we use a HashMap with the code of the vertices (long data type) as keys and vertices as values to access the vertices in the graph since the edge's specifications contain only the codes of the vertices it connects and its cost (distance). In the case where the user gives a point that it is not in our map, we compute a close enough one and use a HashMap with vertex objects as keys and point objects as values to store the original points of the approximated vertices we found. This is done with the aim of returning to the user the same points he or she entered. Finally, we use ArrayLists of vertex objects to store both the points entered by the user and the generated tours.

The reason to use HashMaps is that we do not require our mapping to be stored in order, which allows us to use the most efficient map available in Java.

### 5. ALGORITHM
First, the program creates the city's graph by reading its specifications which are given in two text files, one for the vertices and other for the edges.

Then it reads the URL containing the points from the user, the program finds the nearest vertex to the ones corresponding to the given coordinates, which can be the exact same point. After this, the program will compute the complete subgraph containing only the points of interest. In order to create the subgraph the program will choose between two different algorithms, depending on the number of nodes given by the user: A* algorithm is used if this number is less or equal to five, otherwise it uses Dijkstra's algorithm. For the heuristic of the A* algorithm, it uses the Manhattan distance. Then it will execute one of the following three algorithms, according to user's choice:

1. Natural approximation: the algorithm is a simple sort over the vertices using its 2D coordinates, using a custom comparator to determine whether a point is to the left or to the right of another. This algorithm has two versions: the first (or the fastest) version just performs the sort described above and completes the tour adding the initial vertex, while the second version does the same as the first one, but also computes the subgraph and the reverse tour, compares both total lengths and returns the best tour. Notice that the fastest version does not generate the subgraph. The method for comparing the points is based on [11].

2. Nearest neighbor: This algorithm starts with the first vertex and greedily chooses the nearest neighbor and moves to it. It keeps doing the same thing until all the vertices are visited. Finally, it adds the starting vertex to the end in order to complete the tour. This algorithm will not always give an optimal solution, but it's execution time is much faster than the brute force and can produces better tours than natural approximation.
3. Brute force: we try every possible path starting with the initial vertex, ending at i and passing through all the other vertices different to i. To this path we add the distance from i to the initial vertex to complete the tour.

We now present the complexity analysis for the algorithms where n is the number of points to visit, E the number of edges (335719) and V the number of vertices (179403) in the city's graph.

| | | Wost case complexity | |
|---|---|---|---|
| | | Memory | Time |
| Algorithm | Build subgraph | $O(n^2)$ | $O(n(E+V\log V))$ |
| | Brute Force | $O(n2^n)$ | $O(n^2 2^n)$ |
| | Nearest Neighbor | $O(n)$ | $O(n^2)$ |
| | Natural approximation (Normal mode) | $O(n)$ | $O(n\log n)$ |
| | Natural approximation (Fast mode) | $O(n)$ | $O(n\log n)$ |

Table 1. Complexity table

# 6. IMPLEMENTATION

When starting the program the first thing the user will see is an "Initializing" signal, shown to make the user wait until the program builds the city's graph. After this, the program will show the time required to build the city's graph and ask for the Google Maps URL containing the points to be visited (Figure 1). If the user inputs an invalid URL, the program will tell so and then ask again for a new one (Figure 1). It will repeat this process until the user gives a valid input. If the user wants to end the program from here, he or she may input an 'x' (lower or upper case) as URL and the program will finish.

Figure 1. Starting the program and invalid URL

After this, the program shows a warning telling the user that the route given by our program may differ with Google's because the points we used to build the city's graph are different. Then offers a menu with 6 or 7 different options (Figure 2), depending on how many points were given, to compute a tour which covers all the given points and comes back to the first one (the user should not add the first point at the end of the path).

Figure 2. Program's menu

The first five options are to choose the way the user wants the program to calculate the route, after choosing, the program will show the time needed to generate the route and will take the user to a Google Maps page where he or she can see this route. Since the options number two, three, four and five need to internally create a subgraph, the first time any of these options is chosen, the program will show the time spent creating it.

1. "Natural approximation fast mode" will present a route that tends to be circular and, compared with the other options, it is the fastest, but it will not give any distance and may not show the shortest route.
2. "Natural approximation normal mode" will also present a route that tends to be circular, but this time shows the total distance and may even show a route that is shorter than the one obtained using last mode, although it may still not be the shortest possible route.
3. "Nearest Neighbor" will give a route formed by finding the closest point from the current one, and then move to it. This process will repeat until reaching the last point (corresponding to the first one). It may not give the shortest possible route and its execution time is similar to the second option. This option shows total distance too.
4. "The best of both" will choose the best route between the ones generated by last two options and show the total distance; this will take a little more time. We encourage its usage when there are more than 20 points in the URL and the user wants the shortest route.
5. "Exact" will always show the shortest route with its respective distance. It is potentially much slower than the other options, reason why is not present in the menu when the URL has more than twenty points, since that would take a lot of time. If the user wants, he can expand the maximum of points he or she can input to 'unlimited'. For this, when the program asks for an URL (either when starting the program or changing the current one) "extreme-mode" must be written and press the enter key, then a warning will appear and the program will ask for the URL to continue as usual.
6. "Change URL" lets the user change the current URL, if the new URL is not valid it will indicate so and ask for a new one; if the given input is an 'x' the program will end.
7. "Exit" will end the program.

Something important to consider is that, when the URL contains at least two points that are not connected between them, the program cannot calculate a distance, so it will tell the user so and use the first option ("Natural approximation fast mode") to compute a possible route.

# 7. RESULTS WITH THE MAP OF MEDELLÍN

## 7.1 Execution Time

The following table shows the time in seconds that each algorithm takes to process a route for a different number of vertices. The time required to build the subgraph for that same amount of points is also shown.

|  |  | Vertices | | | |
|---|---|---|---|---|---|
|  |  | 5 | 10 | 15 | 20 |
| Algorithm | Build subgraph | 0,4182 | 1,5146 | 2,1448 | 2,7558 |
|  | Brute Force | 0 | 0,0014 | 0,0428 | 2,1206 |
|  | Best of Both | 0 | 0 | 0,0001 | 0,0002 |
|  | Nearest Neighbor | 0 | 0 | 0 | 0 |
|  | Natural approximation (Normal mode) | 0 | 0 | 0 | 0 |
|  | Natural approximation (Fast mode) | 0 | 0 | 0 | 0 |

Table 2. Execution time (seconds) on a Core i7 2410u processor.

Remember that the Natural Approximation (Fast mode) does not require the subgraph to be computed by the time it is called. For the rest of the algorithms, the program only needs to compute the subgraph once, because after building it, it will be saved and reused until a new subgraph is required (new URL).

As expected, the brute force algorithm is the slower one. For the others, it is hard to compare their running times with such a little amount of vertices.

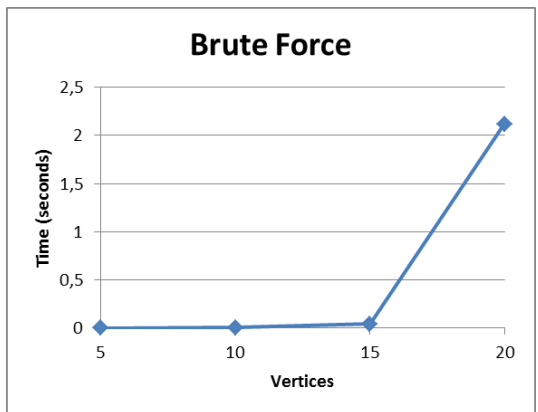

Graphic 1. Brute force's execution time on a Core i7 2410U processor.

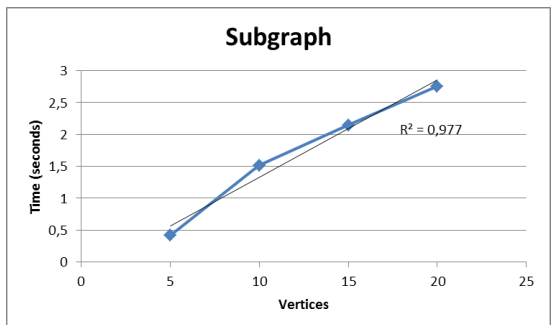

Graphic 2. Subgraph construction time on a Core i7 2410U processor.

Considering current limits imposed by Google Maps in terms of the available amounts of destinations for a single route (ten vertices), it is easy to notice that our algorithm will run in feasible time even when looking for an optimal solution after building the subgraph, and in under two seconds if it has not been build.

## 7.2 Memory space

The following table shows the memory in megabytes that each algorithm takes to process a route for a different number of vertices. Is important to notice that the memory used by the subgraph can only be included in the algorithms that need it.

|  |  | Vertices | | | | |
|---|---|---|---|---|---|---|
|  |  | 5 | 10 | 15 | 20 | 24 |
| Algorithm | Build subgraph | 27 | 58 | 73 | 111 | 136 |
|  | Brute Force | 0 | 1 | 2 | 100 | 4771 |
|  | Nearest Neighbor | 3 | 4 | 3 | 4 | 4 |
|  | Natural approximation (Normal mode) | 3 | 3 | 3 | 4 | 3 |
|  | Natural approximation (Fast mode) | 2 | 4 | 4 | 4 | 4 |

Table 3. Memory used (in MB)

Is easy to notice that for all algorithms, except brute force, the memory is almost constant. This is due, the amount of vertices is not enough to show any difference.

In the other hand, one can see the incredible difference in memory used by the brute force algorithm. This was expected since the memory use increase exponentially ($O(n2^n)$). In graphic 3 we plot the obtained results.

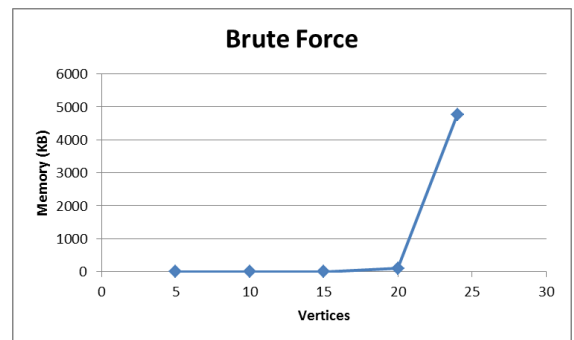

Graphic 3. Memory for Brute force

## 7.3 Total distance

The following graphic shows the total distance (in meters) of the routes found as shortest by the different algorithms for random sets of points.

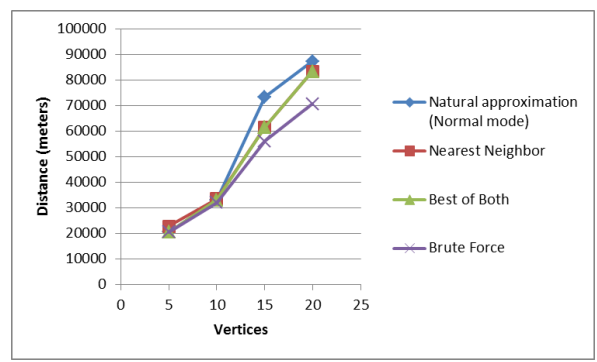

Graphic 4. Total distance

Graphic 4 shows how volatile Nearest Neighbor and Natural Approximation algorithms are: in one case they got the optimal answer whereas in other cases both obtained longer routes than the optimal solution with a difference higher than 10 kilometers.

The code can be found at

https://subversion.assembla.com/svn/planningdeliverysystemmedellin/

## 8. CONCLUSIONS

The problem to calculate a delivery route given some points in a map is not new and has been researched for years. As prove to that, we can find the famous Traveling Salesman Problem or TSP, in which the delivery route may not have repeated points, except for the first one, that is the last one at the same time. This problem was defined in the 1800s and even today, there are not any algorithms that can give the best route efficiently: we still have to choose between efficiency and precision. For the problem of planning delivery routes in Medellín, that can be modeled as the TSP problem, our proposed solution can be of huge help for small companies when their couriers go out for a route because it's unlikely that, on a single trip, they will visit more than 20 points. If the delivery route can consider repeated points it will be harder to solve, so is better to simply solve TSP.

The biggest problem is to find an efficient way to give an answer: algorithms that can give optimal answers require a lot of time and memory, to the point that they cannot be used for big graphs, and algorithms that can give an answer without consuming too much resources, may not be able to give the shortest route. Even if we apply all possible optimizations to the code, it is still not enough to efficiently compute an optimal solution.

Thanks to this work, we were able to understand the limitations a computer has and the need of implementing efficient algorithms with the appropriate data structures, otherwise a program could take too much time to execute, time that the user is not able or willing to spend waiting, since it can get to a point where it needs years to compute an answer. And this gets worse considering that right now we are living in a world were a lot of data is stored, and only accessing it may take a lot of the computer's resources.

## 9. FUTURE WORK

Currently, there are two big limitations we would like to fix in the future:

1. Our graph does not fit Google Map's very well, which makes the distances and the routes to be shown in a very different way in many cases.
2. The only way to add more than 10 points on a route is by working with the URL and the points' latitude and longitude.

## 10. REFERENCES

[1] S. Skiena, The Algorithm Design Manual, New York: Springer, 2008.

[2] E. W. Weisstein, "Eulerian Cycle," Wolfram MathWorld, [Online]. Available: http://mathworld.wolfram.com/EulerianCycle.html. [Accessed 27 August 2016].

[3] E. W. Weisstein, "Eulerian Path," Wolfram MathWord, [Online]. Available: http://mathworld.wolfram.com/EulerianPath.html. [Accessed 28 August 2016].

[4] GeeksForGeeks, "Eulerian path and circuit for undirected graph," GeeksForGeeks, [Online]. Available: http://www.geeksforgeeks.org/eulerian-path-and-circuit/. [Accessed 28 August 2016].

[5] E. W. Weisstein, "Hamiltonian Path," Wolfram MathWorld, [Online]. Available: http://mathworld.wolfram.com/HamiltonianPath.html. [Accessed 28 August 2016].

[6] E. W. Weisstein, "Hamiltonian Cycle," Wolfram MathWorld, [Online]. Available: http://mathworld.wolfram.com/HamiltonianCycle.html. [Accessed 28 August 2016].

[7] E. W. Weisstein, "Graph Cycle," Wolfram MathWorld, [Online]. Available: http://mathworld.wolfram.com/GraphCycle.html. [Accessed 28 August 2016].

[8] GeeksForGeeks, "Backtracking | Set 6 (Hamiltonian Cycle)," GeeksForGeeks, [Online]. Available: http://www.geeksforgeeks.org/backtracking-set-7-hamiltonian-cycle/. [Accessed 28 August 2016].

[9] University of Waterloo, "History of the TSP," University of Waterloo, Enero 2007. [Online]. Available: http://www.math.uwaterloo.ca/tsp/history/index.html. [Accessed 28 August 2016].

[10] A. Levitin, Introduction to The Design and Analysis of Algorithms, 3rd Edition, New Jersey: Pearson, 2012, p. 116.

[11] C. e. dos, "Github," 18 Abril 2016. [Online]. Available: https://github.com/mvpossum/eldiego/blob/master/geometria/orden.radial.cpp. [Accessed 4 September 2016].